# PERANCANGAN TEKNOLOGI CLOUD UNTUK PENJUALAN ONLINE KAIN SONGKET PALEMBANG

**Fikri[1), Leon Andretti Abdillah[2)] Ema Apriyani[3)]**
[1]Program Studi Informatika, Fakultas Ilmu Komputer, Universitas Bina Darma
[2]Program Studi Sistem Informasi, Fakultas Ilmu Komputer, Universitas Bina Darma
[3]Program Studi Komunikasi, Fakultas Ilmu Komunikasi, Universitas Bina Darma
Jl. Ahmad Yani No.12, Palembang, 30264
Telp : +62(711) 515679, Fax : +62(711) 515582
E-mail : *leon.abdillah@yahoo.om

*Abstrak*

*Komputasi awan adalah suatu paradigman dimana informasi secara permanen disimpan di server Internet dan secara temporer disimpan di komputer pengguna (client) termasuk desktop. Kajian ini bertujuan untuk mendesain teknology komputasi awan untuk aplikasi penjualan online guna membantu pengrajin kain songket Palembang dan tidak tertutup kemungkinan untuk usaha kecil mengenah (UKM) lainnyadalam me-manage aset dan pasar bagi produk online mereka.Sehingga ia bisa diakses kapan saja dan dimana saja menggunakan personal computer, laptop, tablet, mobile phone atau smartphone. Infrastruktur komputasi awan yang digunakan dalam kajian ini adalah Software-as-a-service (SaaS) yang memungkinkan pengguna cloud untuk mengeploitasi aplikasi penjualan online tanpa harus menginstal di komputer lokal, menyiapkan server, mempekerjakan operator, biaya perawatan dan sumber daya lainnya sehingga bisa menghemat biaya.*

**Kata kunci:** *Perancangan teknologi cloud, Penjualan online, Kain songket Palembang, UKM, Saas.*

*Abstract*

*Cloud Computing is a paradigm in which information is permanently stored in servers on the Internet and temporarily stored on the user's computer (client) including the desktop. This study aims to design an online sales application based cloud computing technology to help the artisans Palembang songket and do not rule out the possibility for other small medium enterprises (SMEs) to manage assets and market their products online so it can be accessed anytime and anywhere via personal computer, laptop, tabblet, mobile phone or smartphone. Cloud infrastructure used in this study is a Software-as-a-service (SaaS) that enables cloud users to exploit online sales application without having to install on your local computer, set up a dedicated server, operator labor, maintenance costs and other support resources that can make savings in terms of cost.*

## 1. PENDAHULUAN

Teknologi informasi (TI) telah diadopsi oleh beragam aspek kehidupan kita saat ini. kondisi ini terjadi karena TI dapat berkolaborasi dengan banyak bidang pengetahuan lainnya [1]. IT telah membawa perubahan yang fundamental baik untuk organisasi swasta maupun pemerintah [2]. Sehingga TI menjadi suatu *backbone* utama untuk banyak sektor [3]. Beberapa layanan TI yang telah kita nikmati saat ini antara lain: *email* mendistribusikan informasi antara orang-orang dalam suatu organisasi [3], *facebook* sebagai media promosi produk [4] atau kampanye presiden [5], *online storage* bahan pembelajaran [3, 6], dsb, tanpa disadari sebenarnya telah memanfaatkan teknologi *cloud computing*. *Cloud computing* merupakan paradigma baru untuk penempatan (*hosting*) dan pengiriman (*delivering*) melalui *internet* [7]. *Cloud computing* merupakan sebuah mekanisme, dimana sekumpulan IT *resource* yang saling terhubung dan nyaris tanpa batas, baik itu infrastruktur maupun aplikasi dimiliki dan dikelola sepenuhnya oleh pihak ketiga sehingga memungkinkan *customer* untuk menggunakan *resource* tersebut secara *on demand* melalui *network* baik yang sifatnya jaringan *private* maupun *public* [8]. Dengan teknologi *cloud*, pengguna *internet* mulai dari perseorangan, komunitas hingga perusahaan dapat menggunakan aplikasi tanpa harus melakukan instalasi di komputer lokal, mengakses file pribadi mereka di komputer manapun, kapanpun melalui akses *internet*. Keuntungan utama dengan adanya teknologi *cloud* bagi *developers* [9] adalah tidak lagi memerlukan modal yang besar untuk biaya hardware ketika ingin menyebarkan layanan mereka atau biaya *operator*-nya.

*Cloud computing* memiliki tiga kelompok layanan (*services*) [8], yaitu: 1) *Software as a Service* (SaaS), 2) *Platform as a Service (PaaS)*, dan 3) *Infrastructure as a Service (IaaS)*. Pada penelitian ini fokus bahasan lebih kepada *SaaS*. *SaaS* adalah paradigma pengiriman perangkat lunak dimana *software* ditempatkan secara *off-premise* dan disampaikan melalui *web* [10]. Penyedia layanan TI lokal perlu memiliki keahlian di bidang infrastruktur dan *SaaS* [11] untuk merangkul model *cloud* dengan cara mendukung klien mereka dengan transformasi TI mereka ke *Cloud Services* akan menguntungkan. Hal ini tentu saja menguntungkan bagi beberapa perusahaan medium di Indonesia, terutama perusahaan yang bergerak dibidang bukan teknologi informasi (non-IT) seperti: perusahaan manufaktur, distribusi, konsultan, lembaga pendidikan, usaha kecil dan menengah (UKM) dan masih banyak lagi biasanya belum memiliki tenaga IT khusus di perusahaan.

Globalisasi akhir-akhir ini telah menggiring sistem perdagangan dunia menjadi lebih terbuka dan tanpa batas. Bisnis sudah menyatukan dunia sehingga belahan-belahan bumi hanya tinggal berbeda secara koordinat geografis saja. Tahun 2015 era pasar bebas ASEAN (*ASEAN Free Trade Area/AFTA*) segera dimulai. Pada wilayah yang lebih luas lagi, kawasan Asia Pasific juga bersiap menyongsong rezim pasar bebas (*Free Trade Area of the AsiaPacific/FTAAP*) yang lebih keras tantangannya. Negara-negara yang ada di kawasan ini berlomba menyambut era Asia Pasific Trade Area dengan berbagai persiapan dan agenda masing-masing. Indonesia sebagai salah satu negara utama di ASEAN, pencetuk Gerakan Non-Blok (GNB) atau *Non-Aligned Movement* (NAM), sekaligus negara anggota G-16, serta berpenduduk muslim terbesar di dunia juga perlu membekali diri dengan berbagai pengetahuan, teknologi, serta keunggulan lokal agar dapat ikut dalam era tersebut. Dan teknologi terkini yang wajib untuk diterapkan guna mempersiapkan diri di era pasar bebas adalah teknologi berbasis IT. Teknologi ini tidak hanya digunakan oleh perusahaan-perusahaan besar (*enterprise*) namun juga hendaknya dimanfaatkan oleh perusahaan UKM ata small medium enterprises (SMEs). Karena sektor UKM terbukti mampu mendorong kreativitas serta tahan akan goncangan ekonomi berskala besar sehingga lebih mudah dikembangkan dan dikelola. Namun untuk adopsi penggunaan *cloud technology* di Indonesia justru masih di dominasi oleh perusahaan menengah ke atas [12]. Hal ini bisa menjadi tantangan sekaligus peluang untuk penetrasi ke *cluster* perusahaan UKM untuk bisa menggunakan teknologi *cloud* juga.

Pemanfaatan teknologi oleh beberapa perusahaan non-IT hanya bisa diakses secara lokal. Pengguna harus mengakses secara fisik perangkat yang digunakan untuk memperoleh informasi yang dibutuhkan. Padahal seiring dengan perkembangan dunia IT dan semakin pesatnya laju pertumbuhan bisnis perusahaan dibutuhkan informasi secara cepat dan akurat oleh siapapun, kapanpun dan dimanapun. Disinilah perusahaan perlu menerapkan sistem penjualan *online* yang dapat menyajikan informasi-informasi baik bagi pemilik perusahaan maupun bagi para pelanggan yang dapat diakses tanpa terbatas oleh waktu dan tempat melalui komputer, laptop, *tablet*, *smartphone* yang terhubung ke internet.

Songket adalah kain tenun kerajinan yang telah diturunkan oleh dari generasi ke generasi 200 tahun yang lalu di Sumatera [13]. Songket sangat lekat dengan Palembang. Kain songket Palembang adalah salah satu produk yang tidak hanya dikenal secara lokal tetapi juga merupakan salah satu *icon* Palembang di dunia internasional. Maka akan lebih baik jika pemasaran salah satu produk unggulan ini dilakukan secara *online*. Namun Keterbatasan pengetahuan tentang teknologi menjadi salah satu kendala bagi UKM yang berfokus pada pembuatan dan penjualan kain Songket Palembang untuk memasarkan produknya secara *online*. Oleh karena itu solusi bagi permasalahan UKM tesebut ialah sebuah sistem penjualan yang mampu menghasilkan informasi-informasi yang dibutuhkan manajemen secara *up to date*.

Bagian selanjutnya dari artikel ini adalah metode penelitian yang menjelaskan metode pengumpulan data, kebutuhan *hardware* dan *software*, perancangan, topologi, kemudian disusul dengan hasil dan pembahasan dari sistem *cloud* yang dihasilkan. Artikel ini kemudian ditutup dengan simpulan.

## 2. METODE PENELITIAN

### 2.1 Metode Pengumpulan Data

Metode pengumpulan data merupakan suatu pernyataan (*statement*) tentang sifat ,keadaan, kegiatan tertentu dan sejenisnya. Pengumpulan data dilakukan untuk memperoleh informasi yang dibutuhkan dalam rangka mencapai tujuan penelitian [14]. Metode yang digunakan untuk pengumpulan data dalam penelitian ini adalah sebagai berikut: 1) Studi Pustaka: Mengumpulkan data dengan cara mencari dan mempelajari data-data dari buku-buku ataupun referensi lain yang berhubungan dengan penulisan ini. Buku yang yang digunakan penulis sebagai referensi, adapun metode yang digunakan penulis dalam merancang dan mengembangkan dapat dilihat pada daftar pustaka, 2) Observasi: Melakukan pengamatan secara langsung terhadap objek yang akan diteliti secara cermat, 3) Wawancara: Mendapatkan informasi dengan cara bertanya langsung kepada responden dalam hal para Pengerajin Kain Songket Palembang, dan 4) Sampling: Mengumpulkan sampel dokumen, *form*, dan *record* yang *representative* (sampling).

### 2.2 Kebutuhan *Hardware* and *Software*

Dalam penerapannya teknologi yang akan dibangun membutuhkan perangkat keras yang akan diperuntukkkan untuk *server* dan *client* untuk melakukan uji coba sistem. Berikut kebutuhan *hardware* yang digunakan dalam penelitian ini: 1) Komputer *Server Cloud* (Processor Intel EMT64, RAM >2GB, Harddisk >=250GB, NIC), dan 2) Komputer Client (Processor Intel Core, RAM >1GB, Harddisk>=250GB, NIC).

Selain *hardware* teknologi yang akan dibangun membutuhkan *software* sebagai berikut: 1) Sistem Operasi untuk *server cloud* menggunakan *Proxmox VE* (*Virtual Environment*) 3.2 / *Ubuntu Server* 10.4 LTS, 2) Sistem Operasi untuk *client Windows* XP/7/8, 3) Paket *XAMPP for* Linux v1.8.3 (Apache Web Server, PHP *Application Server*, MySQL *Database Server*, *PHPMyAdmin*), dan 4) Aplikasi *inventory* dan *e-commerce* berbasis *web*.

### 2.3 Perancangan

Tahap desain ini akan membuat gambar desain topologi jaringan interkoneksi yang akan dibangun. Pada tahap ini ditetapkan jenis sistem seperti apa yang dapat diterapkan guna memecahkan masalah pada tahap analisis. Sebelum melakukan perancangan, berikut tahapan-tahapan yang direncanakan dalam membangun teknologi *cloud* untuk penjualan *online* diantaranya: 1) Merancang insfrastruktur topologi jaringan yang akan dibangun, 2) Merancang cara kerja aplikasi penjualan bagi para pengrajin kain songket, 3) Menginstal sistem operasi untuk *server cloud*, 4) Konfigurasi *ip address*, dns, *web server* pada sisi *server*, 5) Instalasi aplikasi penjualan *online*, dan 6) Melakukan uji coba system.

### 2.4 Topologi

Perancangan topologi jaringan teknologi *cloud* untuk penjualan *online* kain songket Palembang dalam simulasinya menggunakan topologi *Star* yang disesuaikan dengan proses pembangunan *private cloud* agar semua komputer yang terhubung dalam jaringan dapat melakukan akses ke *server*. Berikut ini topologi jaringan yang dibangun dalam penelitian ini seperti pada gambar 1.

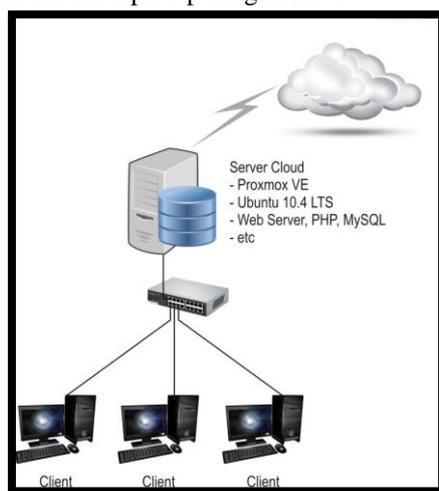

*Gambar 1. Desain Topologi Cloud Online Sales*

Dari topologi di atas dapat digunakan dua skema untuk mendapatkan layanan *cloud*. Pertama akses *cloud* melalui jaringan lokal dimana *server* tidak terhubung ke *internet*. Skema ini hanya digunakan pada tahap uji coba sebelum aplikasi di *upload* ke internet. Skema kedua adalah akses layanan *cloud* melalui jaringan internet. Pada skema ini penulis tidak akan menggunakan *server* lokal sebagai penyedia layanan *cloud* karena semua layanan dalam bentuk aplikasi penjualan *online* akan ditempatkan pada *hosting cloud*.

### 3. HASIL DAN PEMBAHASAN

Setelah beberapa tahap peneliti lakukan mulai dari menyiapkan server, instalasi sistem operasi *server*, menyiapakan *web server*, men-*deploy* aplikasi ke *server* hingga uji coba melalui *client*. Hasilnya ialah dua buah aplikasi yaitu aplikasi *e-commerce* dan aplikasi *inventory* penjualan yang dapat digunakan langsung oleh para pengrajin kain songket Palembang tanpa harus melakukan instalasi dikomputer lokal atau dapat langsung menggunakan aplikasi tersebut melalui komputer, *laptop*, *tablet*, *smartphone* yang terhubung ke *internet*.

### 3.1 Konfigurasi *Hardware*

Dalam penerapannya teknologi yang akan dibangun membutuhkan perangkat keras yang akan diperuntukkkan untuk *server* dan *client* untuk melakukan uji coba sistem. Berikut kebutuhan *hardware* yang digunakan dalam penelitian ini: 1) Komputer *Server Cloud*, terdiri atas: a) Processor Intel EMT64, b) RAM 2 GB (lebih besar lebih baik), c) HDD : 250 GB, d) Additional : 1 NIC, 2) Komputer *Client*, terdiri atas: a) Processor : Intel Core i3 3220, b) RAM : 1 GB, c) HDD : 250 GB, d) Additional : 1 NIC.

### 3.2 Konfigurasi *Software*

Selain hardware teknologi yang akan dibangun membutuhkan *software* sebagai berikut: 1) Sistem Operasi untuk server cloud menggunakan Proxmox VE (Virtual Environment) 3.2 / Ubuntu Server 10.4 LTS, 2) Sistem Operasi untuk client Windows XP/7/8, 3) Paket XAMPP for Linux v1.8.3 (Apache Web Server, PHP Application Server, MySQL Database Server, PHPMyAdmin), dan 4) Aplikasi inventory dan e-commerce berbasis web.

### 3.3 Simulasi

Simlasi disini adalah dimana *requirement* diubah ke dalam sistem yang bekerja (*working system*) yang secara terus menerus diperbaiki. Pada tahap ini sistem yang telah dibangun akan disimulasikan melalui jaringan lokal sebelum diterapkan pada media *online*. Tahapan simulasi dalam penelitian ini adalah: 1) Instalasi *Server* Proxmox VE, 2) Instalasi *Ubuntu Server*, dan 3) Instalasi *Web Server*.

#### 3.3.1 Instalasi Server Proxmox VE

Mesin virtual dalam penelitian ini maksudnya adalah sistem operasi yang dijalankan melalui Proxmox VE. Mesin virtual inilah yang nantinya akan menyediakan layanan bagi aplikasi penjualan online yang akan dibangun. Ada dua pilihan platform untuk membuat mesin virtual ini yaitu berbasis KVM dan OpenVZ. KVM punya keunggulan tersendiri karena mampu menjalankan lebih banyak jenis sistem operasi. Sayangnya tidak semua processor dan mainboard yang ada di pasaran mempunyai fitur virtualisasi yang dibutuhkan untuk menjalankan KVM. Oleh karena itu mesin virtual berbasis OpenVZ bisa menjadi solusi untuk PC yang tidak mendukung KVM.

Sebelum membuat mesin virtual berbasis OpenVZ di Proxmox, kita harus mengunggah template yang dibutuhkan ke server. Sesuaikan juga nama berkas template yang akan diunggah karena jika tidak, Proxmox akan menolak berkas tersebut. Template ini kita unggah menggunakan menu Appliance Template. Seperti gambar 2. Cara lain adalah dengan mengunduh template yang disediakan Proxmox di tab Download. Tentu saja server harus sudah terhubung ke internet seperti gambar 3.

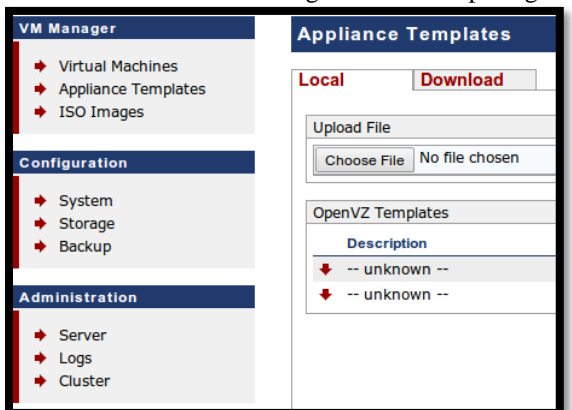
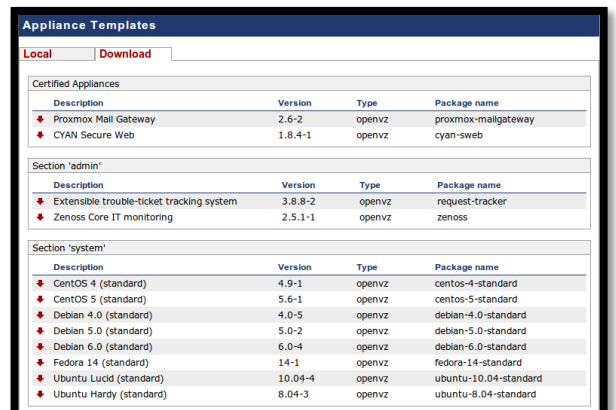

*Gambar 2. Appiance Template*  *Gambar 3. Halaman download Appliance Templates*

#### 3.3.2 Instalasi *Ubuntu Server*

*Web server* merupakan *software* yang memberikan layanan data yang berfungsi menerima permintaan HTTP atau HTTPS dari klien yang dikenal dengan browser web dan mengirimkan kembali hasilnya dalam bentuk halaman-halaman web yang umumnya berbentuk dokumen HTML. Pada bagian ini akan diterangkan cara melakukan instalasi web server di mesin virtual Ubuntu Server 12.04 LTS. Web Server yang digunakan disini adalah LAMP (Linux, Apache, MySQL, PHP). Berikut adalah langkah-langkahnya.

1. Pastikan server terkoneksi dengan internet, jalankan mesin virtual Ubuntu Server 12.04 LTS melalu Proxmox VE. Langkah ini untuk memastikan kita login dengan account root.
2. Langkah awal lakukan instal MySQL. Akan diminta memasukan password MySQL untuk "root".
3. Berikutnya instal Apache Web Server. Lakukan pengujian hasil instalasi web server Apache.
4. Berikutnya instal engine PHP5.
5. Selanjutnya lakukan test untuk memastikan bahwa web server Apache telah berjalan. Selanjutnya beralih ke komputer remote, lalu buka browser kemudian ketikkan http://alamat_ip_mesin_virtual/info.php
6. Berikutnya lakukan instalasi aplikasi untuk manajemen database via web yaitu phpmyadmin. Lakukan pengujian melalu browser dari komputer remote.

### 3.3.3 Instalasi *Web Server*

*Web server* merupakan *software* yang memberikan layanan data yang berfungsi menerima permintaan HTTP atau HTTPS dari klien yang dikenal dengan browser web dan mengirimkan kembali hasilnya dalam bentuk halaman-halaman web yang umumnya berbentuk dokumen HTML. Pada bagian ini akan diterangkan cara melakukan instalasi web server di mesin virtual Ubuntu Server 12.04 LTS. *Web Server* yang digunakan disini adalah LAMP (Linux, Apache, MySQL, PHP).

### 3.4 Implementasi Melalui Jaringan Internet

### 3.4.1 Halaman Utama Web Pengelola

Pada halaman ini calon pelanggan dapat memilih paket yang disediakan oleh pengelola, mendapatkan informasi tentang perusahaan, cara melakukan pemesanan paket, cara melakukan pembayaran dan konfirmasi pembayaran. Berikut ini tampilan halaman utama *web* pengelola (Gambar 4).

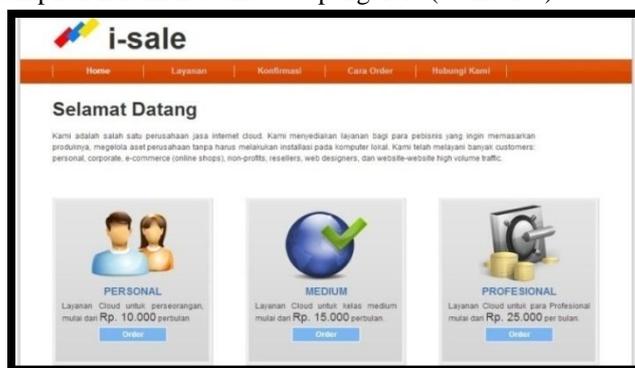

*Gambar 4. Halaman utama web pengelola*

### 3.4.2 Halaman Administrator Pengelola

Untuk melihat siapa saja yang telah melakukan *order* atau pemesanan paket, melihat data konfirmasi, dan mengelola isi *web* yang lain, administrator dapat menggunakan halaman administrator dengan tampilan seperti pada gambar 5.

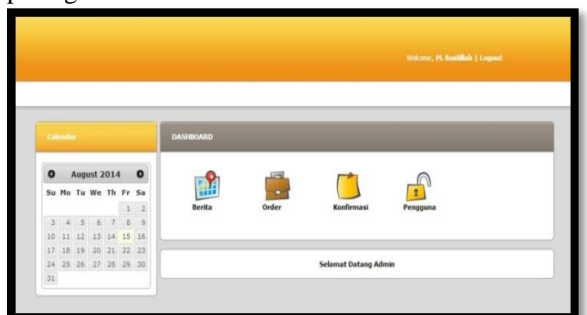

*Gambar 5. Halaman administrator*

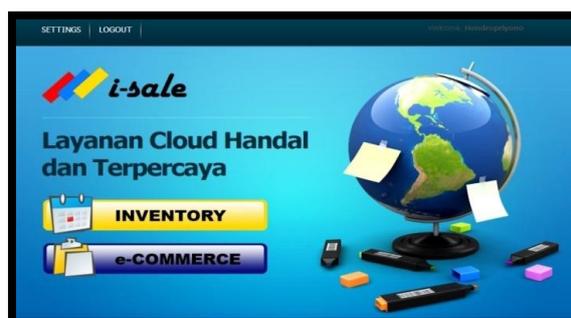

*Gambar 6. Halaman user/pelanggan*

### 3.4.3 Halaman *User* (Pelanggan)

Pada halaman ini (gambar 6) user dapat melakukan data transaksi penjualan dan pembelian, master barang, supplier, membuat laporan dan lain-lain. Pada halaman ini pelanggan dapat membuat toko penjualan online dengan memasukkan kategori barang, jenis barang, menentukan harga.

## 4. SIMPULAN DAN SARAN

Setelah dilakukan uji coba dan sistem penjualan berbasis *cloud*, maka dapat ditarik kesimpulan:
1. Teknologi Cloud Computing memudahkan pengguna mengimplementasikan komputerisasi tanpa mengeluarkan biaya tambahan (perangkat keras dengan spesifikasi tinggi, maintenance, system, dll).
2. Rancangan teknologi Cloud ini dapat juga dimanfaatkan untuk perusahaan lain baik itu untuk perusahaan tingkat menengah ataupun tingkat menengah atas (jasa, dagang, manufaktur, industry).
3. Sistem ini dapat digunakan sebagai media promosi barang melalui layanan e-commerce.

## 5. DAFTAR RUJUKAN